\documentclass[sigconf,screen]{acmart}

\acmConference[HotCarbon'24]{3rd Workshop on Sustainable Computer
Systems}{July 9, 2024}{Santa Cruz, CA}
\acmDOI{}
\acmISBN{}
\setcopyright{rightsretained}

\usepackage[normalem]{ulem}
\usepackage{enumitem}
\usepackage{rotating}
\usepackage{hyperref}
\usepackage{booktabs}
\usepackage{balance}
\usepackage{graphicx}
\usepackage{wrapfig}
\usepackage{algorithm}
\usepackage{algpseudocode}
\usepackage{amsmath}
\usepackage{fancyhdr}
\usepackage{bbm}
\usepackage{multirow}
\usepackage{xcolor,colortbl}
\usepackage{soul}
\usepackage{tikz}
\usepackage{cleveref}
\usepackage{url}
\usepackage{mathtools}
\usepackage{pifont}
\usepackage{makecell}
\usepackage[most]{tcolorbox}

\makeatletter
\def\SOUL@hlpreamble{%
\setul{}{2.2ex}
\let\SOUL@stcolor\SOUL@hlcolor
\SOUL@stpreamble
}
\makeatother
\soulregister\cite7
\soulregister\ref7
\soulregister\eg7
\soulregister\ie7
\soulregister\texttt7

\definecolor{Gray}{gray}{0.85}
\newcolumntype{a}{>{\columncolor{Gray}}c}

\definecolor{myblue}{HTML}{4682B4}

\setlist{noitemsep, leftmargin=*, topsep=0pt, partopsep=0pt}

\crefformat{section}{\S#2#1#3} 
\crefformat{subsection}{\S#2#1#3}
\crefformat{subsubsection}{\S#2#1#3}

\newcommand{\lightred}[1]{\textbf{\textcolor{black!30!red!60!white}{#1}}}
\newcommand{\lightgreen}[1]{\textbf{\textcolor{black!30!yellow!40!green}{#1}}}

\Crefname{figure}{Figure}{Figures}

\newlength{\myeqskip}  
\AtBeginDocument{%
    \setlength\abovedisplayskip{\myeqskip}%
    \setlength\belowdisplayskip{\myeqskip}%
    \setlength\abovedisplayshortskip{\myeqskip-\baselineskip}%
    \setlength\belowdisplayshortskip{\myeqskip}}

\begin{document}

\title{Uncertainty-Aware Decarbonization for Datacenters}



\author{Amy Li}
\affiliation{%
   \institution{University of Waterloo}
   \city{Waterloo}
   \state{ON}
   \country{CAN}}
\email{amy.li2@uwaterloo.ca}

\author{Sihang Liu}
\affiliation{%
   \institution{University of Waterloo}
   \city{Waterloo}
   \state{ON}
   \country{CAN}}
\email{sihangliu@uwaterloo.ca}

\author{Yi Ding}
\affiliation{%
   \institution{Purdue University}
   \city{West Lafayette}
   \state{IN}
   \country{USA}}
\email{yiding@purdue.edu}
\authornote{Corresponding faculty author.}

\begin{abstract}

This paper represents the first effort to quantify uncertainty in carbon intensity forecasting for datacenter decarbonization. We identify and analyze two types of uncertainty—temporal and spatial—and discuss their system implications. To address the temporal dynamics in quantifying uncertainty for carbon intensity forecasting, we introduce a conformal prediction-based framework. 
Evaluation results show that our technique robustly achieves target coverages in uncertainty quantification across various significance levels. We conduct two case studies using production power traces, focusing on temporal and spatial load shifting respectively. The results show that incorporating uncertainty into scheduling decisions can prevent a 5\% and 14\% increase in carbon emissions, respectively. These percentages translate to an absolute reduction of 2.1 and 10.4 tons of carbon emissions in a 20 MW datacenter cluster.

\end{abstract}

\maketitle

\section{Introduction}\label{sec:intro}

Recent years have witnessed an increasing emphasis on decarbonizing datacenters, as datacenters accounted for 2.5--3.7\% of global carbon emissions in 2022~\cite{lavi2022measuring}. This trend is expected to grow due to the escalating demand for computing power driven by machine learning workloads~\cite{wu2022sustainable}.

In this paper, we focus on the Scope 2 carbon emissions~\cite{sotos2015ghg}, which include the indirect carbon emissions associated with the consumption of purchased electricity, steam, heating, and cooling by a company or organization. 
Carbon emissions are a product of the energy consumption and carbon intensity, where the carbon intensity is measured as grams of $CO_2eq$ emitted per $kWh$ of electricity generated or consumed~\cite{maji2024green}. 

Building carbon-free datacenters depends on effective load scheduling, such as suspend-and-resume~\cite{wiesner2021let,radovanovic2022carbon,acun2023carbon} and wait-and-scale~\cite{souza2023ecovisor,hanafy2023carbonscaler}. The core idea of these scheduling strategies is to adapt to renewable energy supplies based on carbon intensity forecasts. Inaccurate carbon intensity forecasts can not only fail to reduce carbon emissions but may even increase them~\cite{cao2023data}. While prior work has introduced various methods for carbon intensity forecasting such as ARIMA models~\cite{bokde2021short} and neural networks~\cite{maji2022dacf,maji2022carboncast}, they focus on point-based estimation, neglecting to account for their uncertainty levels. As prior studies point out, considering uncertainty is crucial for effective scheduling~\cite{wang2023peeling}. In particular, higher uncertainty in predictions prompts conservative load-shifting strategies, whereas lower uncertainty enables more assertive approaches.

To bridge this gap, we tackle the problem of uncertainty quantification of carbon intensity forecasting for datacenter decarbonization. We first identify and analyze two types of uncertainty in carbon intensity forecasting---temporal and spatial---and then illustrate them using the real-world carbon intensity data (\cref{sec:uncertainty}). To address the temporal dynamics in quantifying uncertainty for carbon intensity forecasting, we introduce a conformal prediction-based framework (\cref{sec:method}). 
Evaluation results show that our technique robustly achieves target coverages in uncertainty quantification across various significance levels (\cref{sec:eval}).
We conduct two case studies, each focusing on temporal and spatial load shifting. These case studies are based on the suspend-and-resume scheduling policy~\cite{wiesner2021let,souza2023ecovisor} and use the Google production power trace data~\footnote{These case studies cannot quantify the potential benefits of considering prediction uncertainty for real system implementations. We leave this for future work.}. We summarize our key findings are as follows.
\begin{itemize}
    \item There exist temporal (short-term and long-term) and spatial uncertainty in carbon intensity forecasting.
    \item We demonstrate that even when the point prediction of carbon intensity significantly deviates from the true value, our confidence interval reliably covers the true value.
    \item The case studies on temporal and spatial load shifting demonstrate that incorporating uncertainty into scheduling decisions can prevent a 5\% and 14\% increase in carbon emissions, respectively. Given a 20 MW cluster within a typical datacenter~\cite{varun2020data}, these percentages translate to an absolute reduction of 2.1 and 10.4 tons of carbon emissions. 
\end{itemize}

\section{Uncertainty in Decarbonization}\label{sec:uncertainty}

Decarbonizing datacenters relies on accurate carbon intensity predictions. However, existing predictive tools often exhibit high variations in prediction accuracy, which pose difficulties in decarbonization efforts. These high variations lead to predictive uncertainty, reducing confidence in the predictions and hindering effective decision-making. In this section, we identify and analyze two types of uncertainty in carbon intensity prediction: temporal and spatial. \emph{Temporal uncertainty} refers to the variability of carbon intensity prediction over time. \emph{Spatial uncertainty} refers to the variability carbon intensity prediction across different geographical grids. 

We apply a state-of-the-art carbon intensity prediction method, CarbonCast~\cite{maji2022carboncast}, on real-world carbon intensity data. CarbonCast uses historical energy source mix and weather data to predict hourly carbon intensity for up to 96 hours into the future at one time. We train CarbonCast on 2021 data, validate it on the first half of 2022, and evaluate its performance on the second half of 2022. We compare three regions in the United States: CISO (California ISO), ERCO (Electric Reliability Council of Texas), and ISNE (ISO New England). We use the mean absolute percentage error (MAPE) as the metric for predictability, where lower MAPE value indicates higher prediction accuracy.

\begin{figure}[!t]
	\centering
	\includegraphics[width=0.99\linewidth]{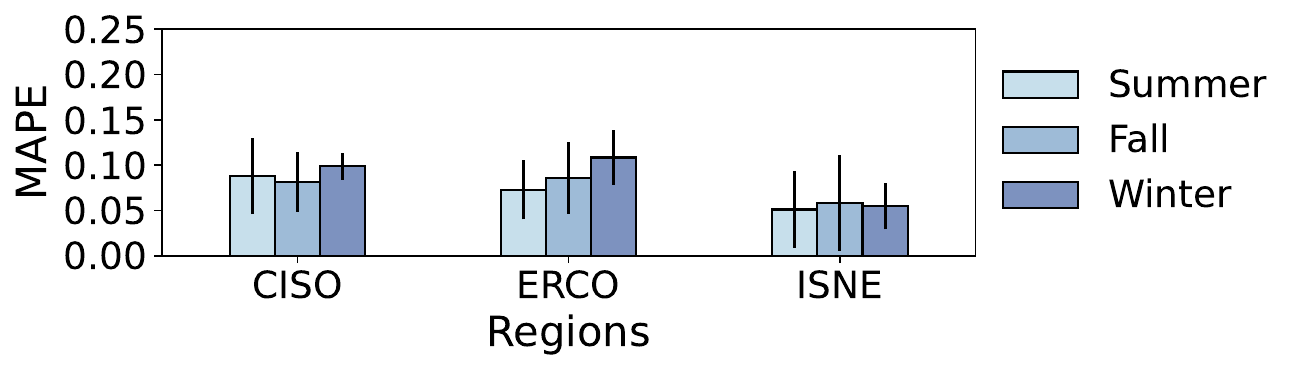} 
  \vspace{-0.1in}
	\caption{Average 24-hour prediction accuracy from July to December in 2022 across three regions. The whiskers indicate standard deviations.}\label{fig:un-agg}
 \vspace{-0.1in}
\end{figure}

\begin{figure}[!t]
	\centering
	\includegraphics[width=0.99\linewidth]{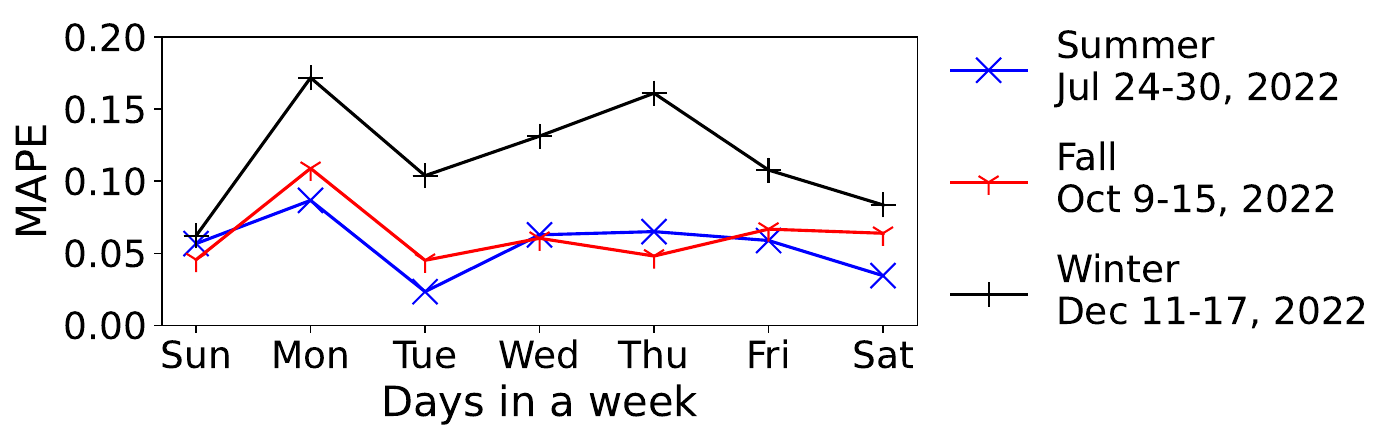} 
  \vspace{-0.1in}
	\caption{Average 24-hour prediction accuracy for a representative week across three seasons in CISO.}\label{fig:un-temporal}
 \vspace{-0.1in}
\end{figure}

\subsection{Temporal Uncertainty}\label{subsec:un-temp}

We characterize temporal uncertainty in short-term and long-term, respectively. CarbonCast predicts up to 24 hours for short-term evaluation and up to 96 hours for long-term evaluation.

\textbf{Short-term.} Figure~\ref{fig:un-agg} shows the average 24-hour prediction accuracy from July to December in 2022 across three regions. This 6-month period is divided into three seasons: summer (July and August), fall (September and October), and winter (November and December). We observe significant seasonal differences in prediction accuracy for all regions, where the best-predicted seasons are fall for CISO, summer for ERCO, and winter for ISNE.

For a finer granular analysis, Figure~\ref{fig:un-temporal} shows the 24-hour prediction accuracy for one representative week during each of the three seasons in CISO in 2022, where the x-axis represents the day of the week. 
We make two observations. First, different seasons exhibit differences in prediction accuracy, with summer and fall displaying 2.1$\times$ and 1.9$\times$ lower MAPEs than winter on average. Second, prediction accuracy fluctuates across days, with summer and fall displaying 3.7$\times$ and 3.2$\times$ lower variances than winter. These results highlight the temporal variability of prediction accuracy from CarbonCast.

\begin{figure}[!t]
	\centering
	\includegraphics[width=0.99\linewidth]{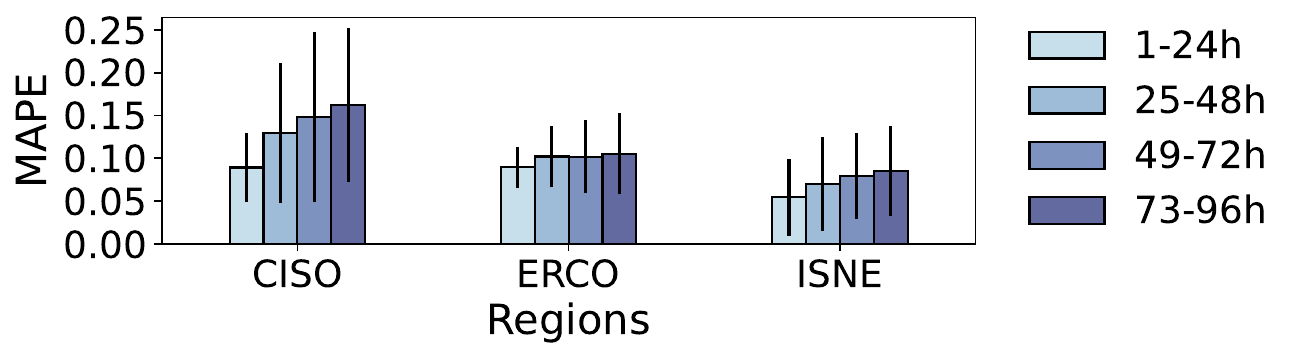}  \vspace{-0.1in}
	\caption{Average prediction accuracy in 4 temporal groups from July to December in 2022 across three regions. The whiskers indicate standard deviations.}\label{fig:un-long}
\vspace{-0.1in}
\end{figure}

\textbf{Long-term.} Besides short-term variations, we observe that prediction accuracy decreases over time. To illustrate such long-term impacts, we let CarbonCast predict 96 hours at one time, and then temporally divide the 96 predictions into 4 groups (i.e., 24 predictions in each group) to compare the prediction accuracy of each group. Figure~\ref{fig:un-long} shows the average 24-hour prediction accuracy of each group from July to December in 2022 across three regions. For all regions, prediction accuracy decreases over time. Specifically, 73--96h predictions have 1.8$\times$, 1.2$\times$, 1.6$\times$ higher MAPEs than 1--24h for CISO, ERCO, and ISNE respectively. Furthermore, CISO's prediction accuracy is highly sensitive to the prediction horizon, while ERCO's is the least affected. This discrepancy may stem from CISO's reliance on renewable energy sources like solar and wind, which are sensitive to weather fluctuations. These results underscore the limitations of CarbonCast's long-term predictability.

\begin{tcolorbox}[colback=white,arc=0pt,outer arc=0pt,colframe=gray,boxrule=1pt,top=0pt,bottom=0pt,left=0pt,right=0pt]
\noindent\textbf{System implications.} Addressing temporal predictive uncertainty in carbon-aware scheduling is critical. A flexible load-shifting policy is essential, enabling dynamic adjustments over time in response to changes in predictive variance. Moreover, it is crucial to recognize that prediction accuracy diminishes with longer horizons. This is especially critical for long-term job scheduling, such as planning days in advance. Neglecting such diminishing prediction accuracy risks higher carbon emissions.
\end{tcolorbox}

\begin{figure}[!t]
	\centering
	\includegraphics[width=0.99\linewidth]{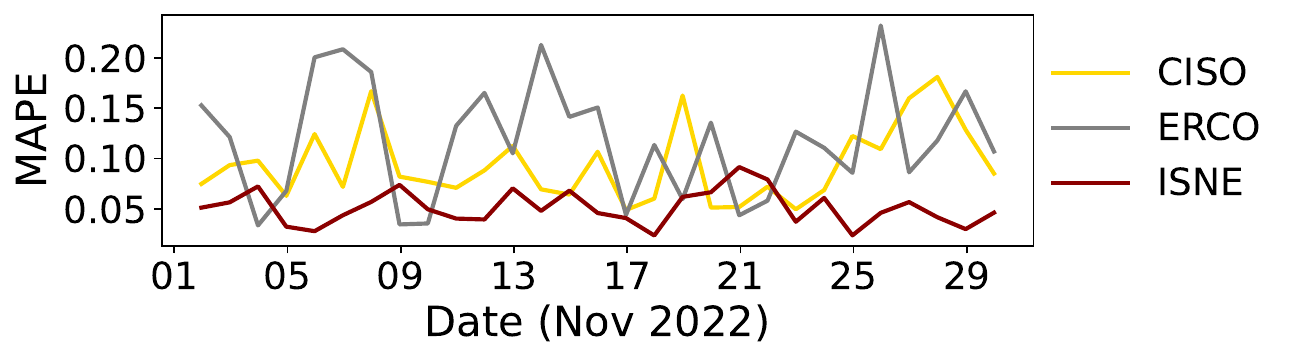}  \vspace{-0.1in}
	\caption{Average 24-hour prediction accuracy for November 2022 in three regions: CISO, ERCO, and ISNE.}\label{fig:un-spatial}
  \vspace{-0.1in}
\end{figure}

\subsection{Spatial Uncertainty}\label{subsec:un-spatial}

Figure~\ref{fig:un-agg} also highlights spatial uncertainty across three regions. It is evident that each region exhibits varying prediction accuracy, where ISNE shows, on average, 1.6$\times$ lower predictability (measured in MAPE) than both CISO and ERCO.

For a finer granular analysis, Figure~\ref{fig:un-spatial} compares the average 24-hour prediction accuracy in November 2022 across three regions, where the x-axis represents the date. The results show that different regions exhibit varying prediction accuracy, with ERCO and ISNE exhibiting 1.5$\times$ and 1.6$\times$ lower MAPEs than CISO on average. Moreover, prediction accuracy fluctuates over different time periods, with ERCO and ISNE showing 3.2$\times$ and 6.4$\times$ lower variance than CISO. These results highlight the spatial variability of prediction accuracy from CarbonCast.

\begin{tcolorbox}[colback=white,arc=0pt,outer arc=0pt,colframe=gray,boxrule=1pt,top=0pt,bottom=0pt,left=0pt,right=0pt]
\noindent\textbf{System implications.} Addressing spatial predictive uncertainty in carbon-aware scheduling is critical. Suppose a long-running workload has two datacenters for execution, A and B, that are located in different regions. 
The carbon intensity is predicted to be low in A at a low confidence, and high in B at a high confidence.
The uncertainty complicates the load migration policy, as it requires assessing whether A's low carbon intensity prediction is robust enough to effectively reduce carbon emissions.
\end{tcolorbox}

\section{Uncertainty Quantification}\label{sec:method}

In this section, we introduce a conformal prediction-based~\cite{shafer2008tutorial} framework for quantifying uncertainty in carbon intensity predictions made by arbitrary prediction algorithms. The fundamental idea is to convert an algorithm's point-based predictions into prediction sets (or a range). Using any pre-trained model, our goal is to generate prediction sets that are guaranteed to contain the true carbon intensity with a user-specified probability. In particular, we train a conformal prediction-based model that starts with CarbonCast to predict the range (also called confidence interval) within which the true carbon intensity value is expected to fall, relative to the CarbonCast's predicted carbon intensity. Sometimes, this model may determine that the CarbonCast prediction is highly ``non-conformal" and therefore provides a confidence interval that is likely to include the true carbon intensity value, deviating significantly from the CarbonCast prediction.


The problem setup is as follows. Consider a sequence of observations $(x_t, y_t), t=1,2,...$, where $x_t\in\mathbb{R}^d$ denotes the features such as energy production and weather, and $y_t$ represents the corresponding true carbon intensity. Let the first $T$ observations $\{(x_t, y_t)\}_{t=1}^T$ be the training data. Our goal is to construct the confidence intervals $\hat{C}_{t-1}(x_t)$~\footnote{The subscript $t-1$ indicates the interval is constructed using previous up to $t-1$ many observations. } sequentially from $T+1$ such that $\hat{C}_{t-1}(x_t)$ will contain the true carbon intensity values with a high probability $1-\alpha$ while the confidence interval is as narrow as possible. 
\begin{align}\label{eq:prob}
 \mathbb{P}\left(y_t\in \hat{C}_{t-1}(x_t) \right) \geq 1-\alpha, \forall t.
\end{align}
$\hat{C}_{t-1}(x_t)$ depend on $\alpha$ and point predictions $\hat{y}_t:=\hat{f}(x_t)$, where $\hat{f}$ is any predictive model (CarbonCast in this case).

We address this problem using conformal prediction. The key ingredient of conformal prediction is the non-conformity scores, which help us evaluate how ``unusual" a new prediction is compared to the predictions made from the calibration data~\footnote{The calibration data, also the validation data in this context, is a subset extracted from the training data and used to estimate the confidence levels of the predictions.}. These scores determine the distance between new predictions and the set of previous observations, which were used as a reference. Essentially, the more a new prediction deviates from the previous data, the less ``conformal" it is, resulting in a higher nonconformity score. A commonly used nonconformity score is the prediction residual:
\begin{align}\label{eq:residual}
    \hat{\epsilon}_t = y_t - \hat{y}_t.
\end{align}
We calculate the nonconformity scores on the calibration data, and then sort them in a descending order to obtain a sorted residual list $\mathcal{E}_t^T$. Then, the confidence interval with $1-\alpha$ probability that satisfies \Cref{eq:prob} will be
\begin{align}
[\hat{y}_t + q_{\alpha/2}(\mathcal{E}_t^T)~,~~\hat{y}_t + q_{1-\alpha/2}(\mathcal{E}_t^T)],    
\end{align}
where $q_{1-\alpha}$ is the $1-\alpha$ quantile function over the set of sorted residuals.

This is the procedure of conventional conformal prediction~\cite{shafer2008tutorial}. However, quantifying uncertainty for carbon intensity prediction is more challenging due to the temporal dynamics in time-series data. As more grids increasingly integrate renewable energy sources, the distribution of carbon intensity will shift. Therefore, we want the conformal prediction method to account for dependencies between data points over time. To address this, we leverage the sequentially predictive conformal interval (SPCI) algorithm~\cite{xu2023sequential}. The key steps are outlined in \Cref{alg}.

\begin{figure}[!t]
	\centering
 \vspace{-0.1in}
	\includegraphics[width=0.99\linewidth]{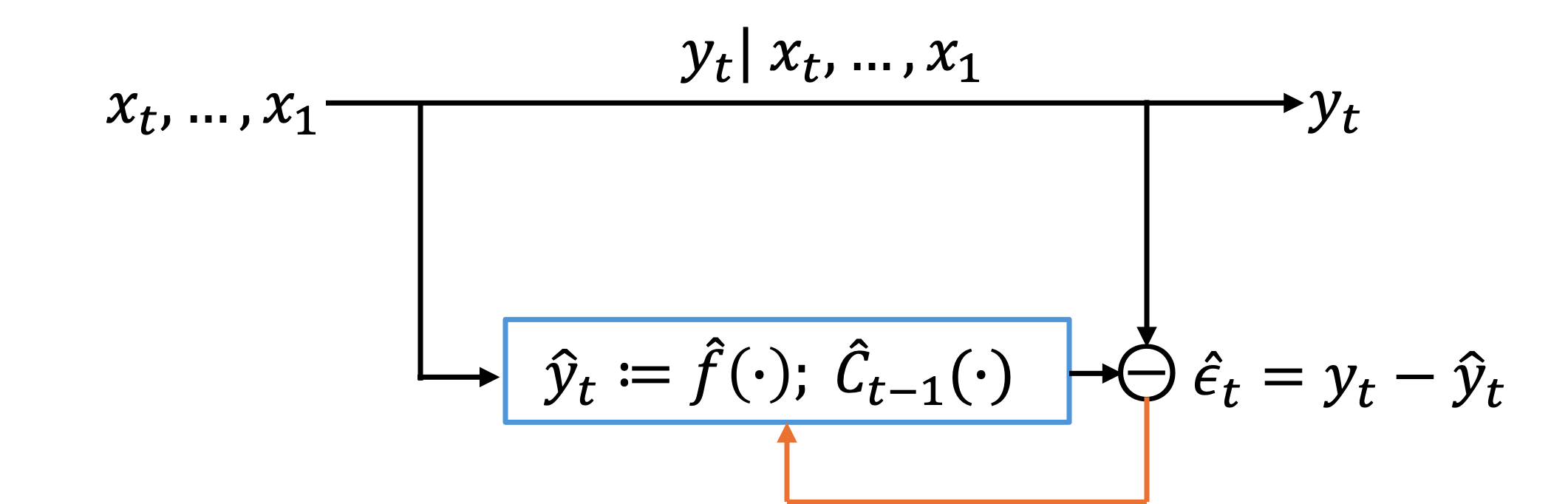} 
  \vspace{-0.1in}
	\caption{The feedback mechanism (the red arrow), where $\hat{C}_{t-1}(x_t)$ is updated based on the residuals at each step.}\label{fig:feedback}
 \vspace{-0.2in}
\end{figure}

The novelty in the SPCI algorithm is the feedback mechanism illustrated in Figure~\ref{fig:feedback}, which encodes temporal dependence information in the prediction residuals. Specifically, $\hat{C}_{t-1}(x_t)$ is updated based on the updated residuals obtained at each step. Additionally, instead of directly using empirical prediction residuals, SPCI trains quantile random forest models autoregressively to predict the conditional quantiles of future unobserved residuals to formulate the residual list. This further accounts for the temporal dependencies between data points over time.

\begin{algorithm}[!t]
	\small
	\caption{SPCI for Uncertainty Quantification}\label{alg}
	\begin{algorithmic}[1]
		\Require $\{(x_t, y_t)\}_{t=1}^T$ \Comment{Training data} 
        \Require $\mathcal{A}$ \Comment{Carbon intensity forecast algorithm (e.g., CarbonCast~\cite{maji2022carboncast})} 
        \Require $\alpha$ \Comment{Significance level}
        \Ensure $\hat{C}_{t-1}(x_t),~t>T$ \Comment{Confidence intervals}
		\State Obtain $\hat{f}$ and residual set $\{\hat{\epsilon}\}_{t=1}^T$ with $\mathcal{A}$ and $\{(x_t, y_t)\}_{t=1}^T$.
		\For {$t > T$}
        \State Obtain $\hat{C}_{t-1}(x_t)$ as in the SPCI algorithm~\cite{xu2023sequential}.
		\State Obtain new residual $\hat{\epsilon}_t$.
  \State Add $\hat{\epsilon}_t$ to the residual set and remove the oldest residual.
		\EndFor
	\end{algorithmic}
\end{algorithm}

\section{Evaluation} \label{sec:eval}

\begin{table*}[!t]
\begin{center}
\caption{Coverage results for three regions over six months at three significance levels. The Coverage column in gray shade shows the overall coverage results, indicating the proportion of SPCI's confidence intervals (CIs) that cover the true values. We further break down the results into two categories: CIs that cover true values ($T_{\rm covered}$) and those that do not ($T_{\rm uncovered}$). Within each category, we also differentiate between CIs that cover CarbonCast predictions ($P_{\rm covered}$) and those that do not ($P_{\rm uncovered}$). } \vspace{-0.1in}
\label{tbl:coverage}
\setlength{\tabcolsep}{1.5pt}
\small
\begin{tabular}{l|a|ll|ll|a|ll|ll|a|ll|ll} \toprule
\multicolumn{1}{c}{} & \multicolumn{5}{c}{$\alpha=0.1$}  & \multicolumn{5}{c}{$\alpha=0.05$}    & \multicolumn{5}{c}{$\alpha=0.01$}                          \\ \midrule
 \multicolumn{1}{c}{} & \multicolumn{1}{c}{}  & \multicolumn{2}{c}{$T_{\rm covered}$} & \multicolumn{2}{c}{$T_{\rm uncovered}$} & \multicolumn{1}{c}{}  & \multicolumn{2}{c}{$T_{\rm covered}$} & \multicolumn{2}{c}{$T_{\rm uncovered}$} & \multicolumn{1}{c}{} & \multicolumn{2}{c}{$T_{\rm covered}$} & \multicolumn{2}{c}{$T_{\rm uncovered}$}\\ \midrule
& Coverage  & $P_{\rm covered}$   & $P_{\rm uncovered}$    & $P_{\rm covered}$        & $P_{\rm uncovered}$        & Coverage & $P_{\rm covered}$   & $P_{\rm uncovered}$    & $P_{\rm covered}$        & $P_{\rm uncovered}$        & Coverage & $P_{\rm covered}$   & $P_{\rm uncovered}$   & $P_{\rm covered}$        & $P_{\rm uncovered}$        \\ \midrule
CISO & 92.41   & 81.94 & 10.47 & 6.24       & 1.35       & 96.34   & 93.62 & 2.72  & 3.49       & 0.16       & 99.28   & 99.28 & 0     & 0.72       & 0          \\
ERCO & 92.02   & 70.44 & 21.58 & 4.7        & 3.28       & 96.02   & 76.75 & 19.27 & 2.28       & 1.7        & 99.09   & 91.62 & 7.47  & 0.77       & 0.14       \\
ISNE & 90.92   & 53.49 & 37.43 & 4.93       & 4.14       & 95.74   & 67.85 & 27.89 & 2.68       & 1.58       & 98.93   & 86.57 & 12.36 & 0.74       & 0.33 \\ \bottomrule    
\end{tabular}
\end{center}
\end{table*}


\begin{figure*}[!t]
	\centering
	\includegraphics[width=0.99\linewidth]{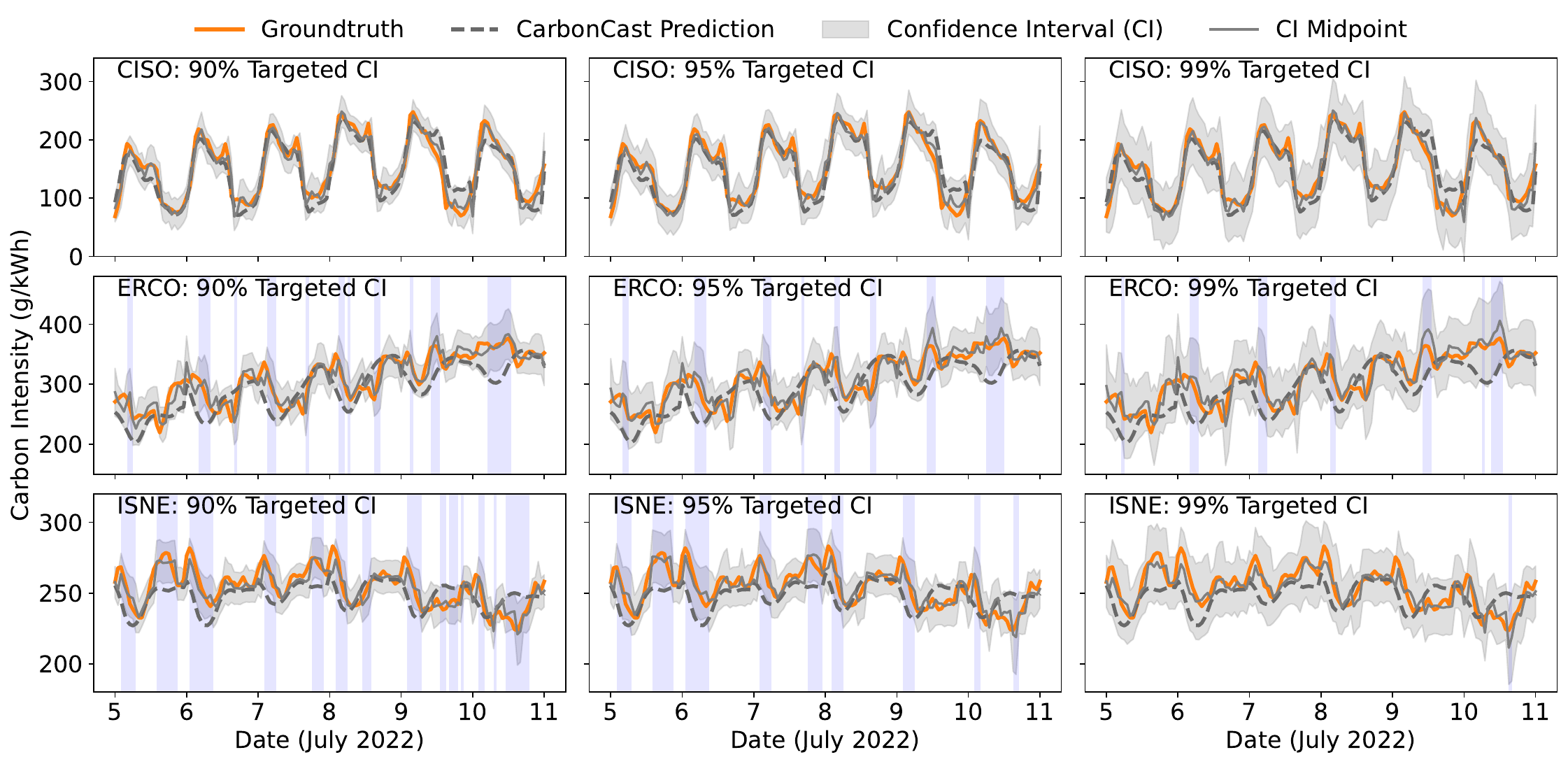}  \vspace{-0.1in}
	\caption{Confidence intervals across one week for three regions at three significance levels $\alpha$ = 0.1, 0.05, and 0.01. The light blue shaded areas indicate the times when the true carbon intensity values are covered but CarbonCast predictions are not.}\label{fig:ci-all}
\end{figure*}

We evaluate our approach from two aspects: uncertainty quantification on real-world carbon intensity data and simulated carbon emissions in case studies for temporal and spatial load shifting.

\subsection{Uncertainty Quantification}\label{subsec:uncertainty_quantification}

\textbf{Evaluation methodology.} Same as~\Cref{sec:uncertainty}, we examine three regions: CISO, ERCO, and ISNE. We collect the historical energy source data from EIA~\cite{eia}, the 96-hour weather forecasts data from NCEP GPS ds084.1~\cite{cisl_rda_ds084.1}, and day-ahead solar/wind forecasts for CISO from OASIS~\cite{caiso}. All data are processed at hourly intervals. We compute the average grid carbon intensity based on the weighted average of carbon emitted by each source~\cite{maji2022carboncast}. 

We apply our uncertainty quantification technique to a state-of-the-art carbon intensity prediction tool CarbonCast~\cite{maji2022carboncast}. We run CarbonCast to get the hourly carbon intensity predictions. The CarbonCast predictions and the ground truth carbon intensity data are then passed to the SPCI framework to obtain confidence intervals. Both CarbonCast and SPCI train on 2021 data, validate/calibrate on the first half of 2022, and test on the second half of 2022. We use \emph{coverage} to evaluate uncertainty quantification, which is the proportion of times that an hourly point estimate's predicted confidence interval (CI) contains the true carbon intensity value. We focus on three significance levels:  $\alpha=$0.1, 0.05, or 0.01, corresponding to targeted coverages of 90\%, 95\%, and 99\%, respectively. 

\textbf{Results.} Table~\ref{tbl:coverage} summarizes the coverage results, both aggregated and breakdown, for three regions over six months at three significance levels. The targeted coverage levels are met across all regions, with the exception of ISNE at $\alpha=0.01$, where the coverage slightly falls short of the expected 99\%. Notably, CISO consistently exhibits higher coverage compared to ERCO and ISNE across various $\alpha$. This discrepancy can be attributed to two factors. Firstly, CISO maintains a greater reliance on renewable energy production than the other regions. Secondly, CISO enhances prediction accuracy by incorporating additional solar and wind inputs. These findings underscore the efficacy of our approach in quantifying uncertainty for carbon intensity forecasting.

In Table~\ref{tbl:coverage}, we further break down the coverage results into two categories: CIs that cover true values ($T_{\rm covered}$) and those that do not ($T_{\rm uncovered}$). In each category, we differentiate between CIs that cover CarbonCast predictions ($P_{\rm covered}$) and those that do not ($P_{\rm uncovered}$). We observe that when CIs cover the true values, they often also cover CarbonCast predictions. However, sometimes the CIs cover only the true values and miss CarbonCast predictions. This outcome aligns with our goal, which is to enhance the coverage of true carbon intensity values, rather than the predictions. 

To provide a detailed view of the coverage results, \Cref{fig:ci-all} presents hourly true, CarbonCast predictions, SPCI's confidence intervals (CIs), and the midpoints (the points that lie in the middle of the CI.) of the CIs at various significance levels, spanning a week. Our observations are as follows.
\begin{itemize}
    \item As $\alpha$ increases, the fractions of points---groundtruth, CarbonCast predictions, and CI midpoints---falling outside of the CIs decrease. This is expected, as higher $\alpha$ values result in wider CIs.    
    \item Even when CarbonCast predictions deviate significantly from the true values, such as ERCO's 90\% targeted CI on July 10th, 2022, our CIs still cover the true values (indicated by the blue shaded areas). This is evidenced by the fact that the midpoints of our CIs are closer to the true values than the CarbonCast predictions.   
    \item When CarbonCast predictions deviate significantly from the true values, the CIs become wider (as seen with ERCO and ISNE between July 10th and 11th). This is useful for decision-making, as wider CIs indicate lower confidence levels and, consequently, conservative scheduling decisions.
    \item CISO consistently exhibits higher coverage and narrower CIs compared to ERCO and ISNE. This could be attributed to CISO's more consistent carbon intensity patterns from day to day, facilitating more accurate predictions.
\end{itemize}

\subsection{Case Studies for Load Shifting} \label{subsec:case-studies}

\textbf{Evaluation methodology.} We simulate load shifting using power traces from Google production systems~\cite{varun2020data}. Specifically, we take the power data from one cluster and apply it to different regions and times for comparative studies. In our scenario, we assume the workload is executed on a cluster with a peak power of 20 MW, based on actual power data from a Google production data center trace. We then compare the reduction in carbon emissions by accounting for the uncertainty of carbon intensities when making load-shifting decisions. Because the power trace data are normalized by Google, our simulated results are also presented as normalized carbon emissions. We would like to clarify that the case studies in this section serve as proof of concept to demonstrate two key points for load shifting decision makers: (1) they should consider both predicted carbon intensity values and their associated uncertainty levels, and (2) they should shift load only when the confidence in the predictions is sufficiently high. These case studies are not real system implementations, which would be far more complex and need to consider additional system-wide factors not addressed here.

We use the widely recognized and effective scheduling policy, suspend-and-resume (also called WaitAWhile), for temporal and spatial load shifting~\cite{wiesner2021let,souza2023ecovisor}. The idea is to suspend work at times or in regions with higher predicted carbon intensity and resume work at times or in regions with lower predicted carbon intensity. Rather than introducing a new scheduling algorithm, we apply the existing suspend-and-resume scheduling algorithm to the prediction results in our case studies. This scheduling algorithm serves our purpose by demonstrating that effective load shifting should consider both predictions and their associated uncertainty levels.

\begin{table}[!t]
\caption{Aggregated temporal shifting results over six months across three regions. Misleading Predictions represents the proportion of days when the predicted carbon intensity for the current day is lower than that of the next day, while in reality, the opposite is true. Increased Emissions represent the proportion of increased carbon emissions if shifting load from the current day to the next day in those cases. }\vspace{-0.1in}
\label{tbl:tbl-temp-agg}
\small
\begin{center}
\begin{tabular}{llll} \toprule
     & CISO             & ERCO                & ISNE  \\ \midrule
Misleading Predictions  & 16.8\% & 10.6\% & 13.4\%  \\
Increased Emissions & 4.3\% & 6.6\%&  4.6\% \\    \bottomrule                    
\end{tabular}
\end{center} 
\vspace{-0.1in}
\end{table}

\begin{figure}[!t]
	\centering
	\includegraphics[width=1\linewidth]{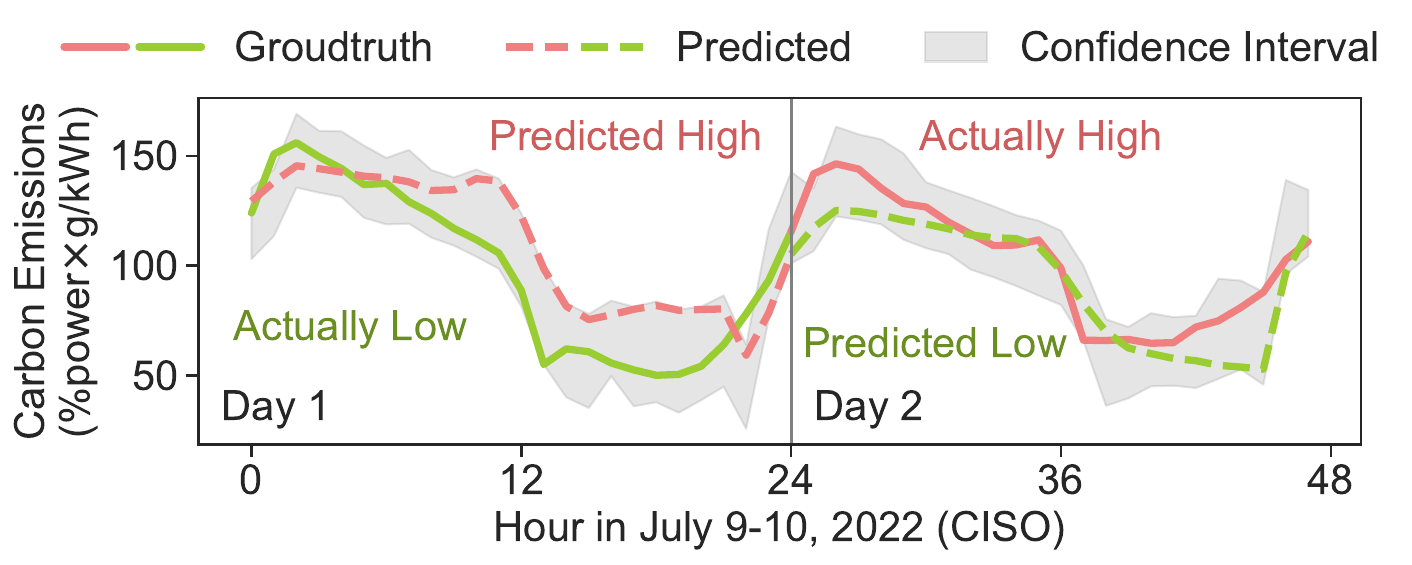}  \vspace{-0.2in}
	\caption{In temporal load shifting, predicted carbon emissions are higher on Day~1 than on Day~2, but true emissions show the opposite trend, with their confidence intervals being roughly similar. Scheduling solely based on predictions would result in a 5\% increase in carbon emissions. The high and low carbon emissions in prediction/groundtruth are marked in \lightred{red} and \lightgreen{green}, respectively.}\label{fig:case-temp}
 \vspace{-0.1in}
\end{figure}


\begin{table}[!t]
\begin{center}
\caption{True and predicted carbon emissions with 90\% confidence intervals per day for CISO. The results are normalized based on the groundtruth value of Day 1.}
\label{tbl:tbl-temp}
\vspace{-0.1in}
\small
\begin{tabular}{llll} \toprule
     & Groundtruth              & Predicted                & Confidence Interval  \\ \midrule
Day 1 & \lightgreen{1.00}        & \lightred{1.13}      & $[0.83, 1.21]$     \\
Day 2 & \lightred{1.05}       & \lightgreen{0.96}      & $[0.84, 1.20]$   \\  \bottomrule                    
\end{tabular}
\end{center} 
\vspace{-0.1in}
\end{table}

\textbf{Temporal load shifting.} In simulating temporal load shifting, we predict carbon intensity for two consecutive days. We then apply power data to obtain their predicted and true carbon emissions. Table~\ref{tbl:tbl-temp-agg} summarizes the aggregated results over six months across three regions, which includes (1) the proportion of days when the predicted carbon intensity for the current day is lower than that of the next day, while in reality, the opposite is true; and (2) the proportion of increased carbon emissions if shifting load from the current day to the next day. Across all regions, 10.6--16.8\% of times show that the predicted carbon intensity for two consecutive days exhibits an opposite trend compared to their true values. If load shifting is performed based solely on these point predictions, it could result in a 4.3--6.6\% increase in carbon emissions. These results indicate that making load-shifting decisions based solely on point carbon intensity predictions is unreliable. Next, we will illustrate how incorporating uncertainty levels of predictions can inform better decision-making using a two-day simulation result. 

\Cref{fig:case-temp} shows the hourly normalized carbon emission results for two days, while \Cref{tbl:tbl-temp} summarizes the total carbon emission results aggregated for each day. We can see that Day~2 shows lower predicted total carbon emissions than Day~1, yet significantly higher true total carbon emissions than Day~1. If workload scheduling to Day~2 is solely based on predicted carbon intensity and emissions, it could result in a 5\% increase in total carbon emissions.
For a cluster that has a 20 MW power in a datacenter~\cite{varun2020data}, this increase can lead to \emph{2.1 tons} of extra CO\textsubscript{2}e. 
However, considering the confidence interval reveals that Day 2 and Day 1 have very similar confidence intervals. Hence, scheduling to Day 2 does not guarantee clear benefits over Day 1. This case study underscores the importance of considering confidence intervals for effective temporal load shifting.

\begin{table}[!t]
\begin{center}
\caption{Aggregated spatial shifting results over six months across three regions. Misleading Predictions represents the proportion of days when the predicted carbon intensity for the target region is lower than that of the source, while in reality, the opposite is true. Increased Emissions represent the proportion of increased carbon emissions if shifting load from the source region to the target in those cases. }
\label{tbl:tbl-spatial-agg}
\small
\vspace{-0.1in}
\begin{tabular}{llll} \toprule
Source & Target & Misleading Predictions & Increased Emissions \\ \midrule
\multirow{2}{*}{CISO}   & ERCO   & 5.0\%          & 3.1\%       \\
   & ISNE   & 7.8\%          & 5.8\%       \\ \hline
\multirow{2}{*}{ERCO}    & CISO   & 2.2\%          & 2.7\%       \\
   & ISNE   & 5.0\%          & 3.5\%       \\ \hline
\multirow{2}{*}{ISNE}    & CISO   & 4.5\%          & 4.3\%       \\
   & ERCO   & 2.8\%          & 7.3\%      \\ \bottomrule                    
\end{tabular}
\end{center} 
\vspace{-0.1in}
\end{table}

\begin{figure}[!t]
	\centering
	\includegraphics[width=1\linewidth]{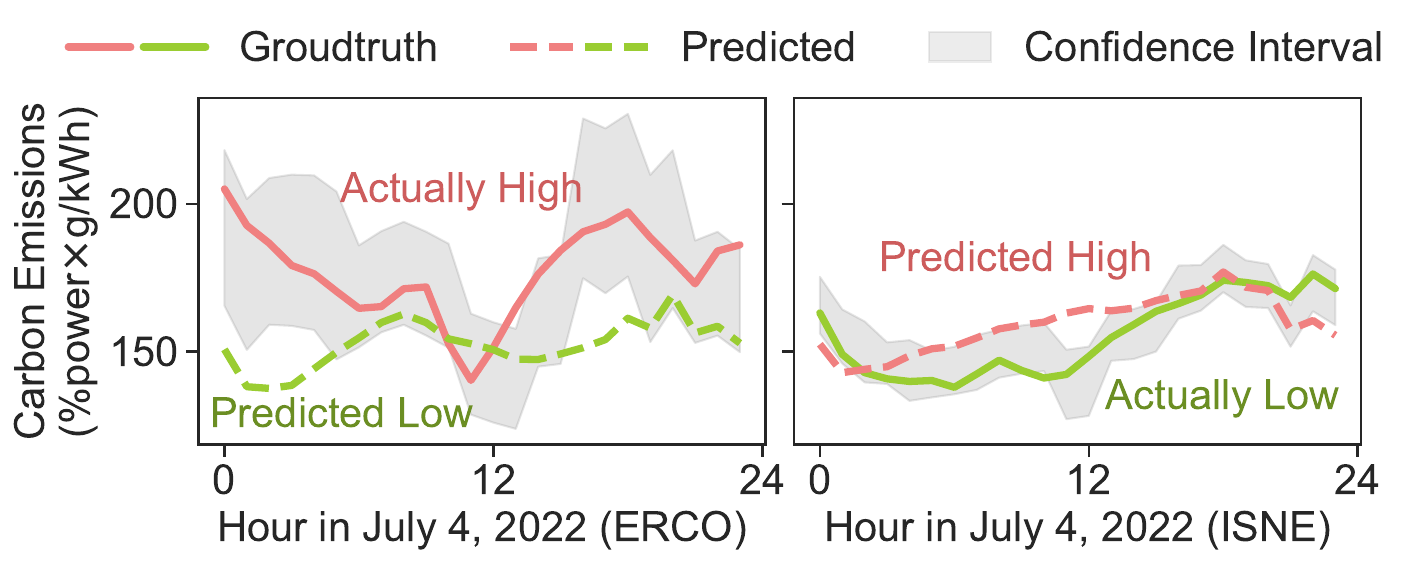}  \vspace{-0.1in}
	\caption{In spatial load shifting, predicted carbon emissions are higher in ISNE than ERCO on the same day, but true emissions show the opposite trend. ERCO's confidence intervals are much wider than ISNE's. Scheduling solely based on predictions would result in a 14\% increase in carbon emissions. The high and low carbon emissions in prediction/groundtruth are marked in \lightred{red} and \lightgreen{green}, respectively.} \label{fig:case-spatial}
\end{figure}

\begin{table}[!t]
\caption{True and predicted carbon emissions in a day with 90\% confidence intervals for ERCO and ISNE. The results are normalized based on ERCO's groundtruth value.}\vspace{-0.1in}
\label{tbl:tbl-spatial}
\small
\begin{center}
\begin{tabular}{llll} \toprule
     & Groundtruth              & Predicted                & Confidence Interval  \\ \midrule
ERCO & \lightred{1.00}& \lightgreen{0.86} & $[0.86, 1.11]$  \\
ISNE & \lightgreen{0.87} & \lightred{0.90} &  $[0.83, 0.93]$ \\    \bottomrule 
\end{tabular}
\end{center} 
\vspace{-0.1in}
\end{table}

\textbf{Spatial load shifting.} In simulating spatial load shifting, we predict carbon intensity for two regions---source (current region) and target (potential region to shift)---on the same day. We then apply power data to calculate their predicted and true carbon emissions. Table~\ref{tbl:tbl-spatial-agg} summarizes the aggregated results over six months, with each case involving a source and target grid. Across all cases, 2.2--7.8\% of times show that the predicted carbon intensity for two regions exhibits an opposite trend compared to their true values. If load shifting is performed based solely on point predictions, it could result in a 2.7--7.3\% increase in carbon emissions. These results indicate that making load-shifting decisions based solely on point predictions is unreliable. Next, we will illustrate how incorporating uncertainty levels of predictions can inform better decision-making using a two-region simulation result on a single day.  

\Cref{fig:case-spatial} shows the hourly normalized carbon emission results for spatial load shifting between ERCO and ISNE on the same day, while \Cref{tbl:tbl-spatial} summarizes the total carbon emission results aggregated over 24 hours. We can see that ERCO shows lower predicted total carbon emissions than ISNE, yet significantly higher true total carbon emissions than ISNE. If workload scheduling to ERCO is solely based on a point estimation of carbon intensity, it could result in a 14\% increase in total carbon emissions. 
Like the previous case study, given a 20~MW datacenter cluster \cite{varun2020data}, this increase means an extra 10.4 tons of CO\textsubscript{2}e. 
However, considering the confidence interval reveals ERCO's wider confidence intervals, with its lower bound not surpassing ISNE's upper bound. Hence, scheduling to ERCO does not guarantee clear benefits over ISNE. This case study underscores the importance of considering confidence intervals for effective spatial load shifting.
\section{Conclusion}

Decarbonizing datacenters demands accurate carbon intensity predictions and uncertainty levels. This study pioneers quantifying such uncertainty and highlights its significance in carbon-aware scheduling. Our evaluation of real-world carbon intensity and power data demonstrates the effectiveness of our technique. We hope this work can inspire system researchers to consider uncertainty when designing future sustainable computing systems.

\section*{Acknowledgements}

We thank Tom Anderson for both patience and the detailed feedback that greatly improved this final version of the paper. We also thank the anonymous reviewers for their helpful feedback. 
This work is supported by the Natural Sciences and Engineering Research Council of Canada (NSERC) and the Undergraduate Research Assistantship (URA) program of the Cheriton School of Computer Science at the University of Waterloo.

\bibliographystyle{plain}
\bibliography{reference}

\end{document}